\documentclass[11pt]{article}
\usepackage[margin=1in]{geometry}
\usepackage[T1]{fontenc}
\usepackage{mathptmx}
\usepackage{cite}
\usepackage{amsmath,amssymb,amsfonts}
\usepackage{graphicx}
\usepackage{textcomp}
\usepackage{booktabs}
\usepackage{array}
\usepackage{placeins}
\usepackage{authblk}
\usepackage[hidelinks]{hyperref}
\usepackage[font=small,labelfont=bf]{caption}
\newcommand{\ket}[1]{|#1\rangle}
\newcommand{\bra}[1]{\langle#1|}
\newcommand{\pout}{p_{\mathrm{out}}}
\newcommand{\area}{\mathcal{A}}

\title{Direct-Coupled Architectures for Magic-State Factories}
\author[1]{Sungyeon Kook}
\author[2]{Yujin Kang}
\author[1]{Ilkwon Sohn}
\author[2]{Jun Heo\thanks{Corresponding author: junheo@korea.ac.kr}}
\affil[1]{Korea Institute of Science and Technology Information, Daejeon, Republic of Korea}
\affil[2]{Korea University, Seoul, Republic of Korea}
\date{}

\begin{document}
\maketitle

\begin{abstract}
In surface-code magic-state factories, the interface that passes level-1 T states to level~2, i.e., the transport channels, waiting spots, and $\ket{0}$ correction patches, occupies a large part of the factory area. We remove this interface by applying adaptive measurement, a principle known from gate teleportation, to the consumption of T states: in adaptive X/Y consumption, the T-state patch used for a $\pi/8$ rotation is read out in the X or Y basis depending on the merge outcome. The basis choice itself performs the $\pi/4$ correction, so no correction patch is needed and a T state can be consumed wherever it already sits. Building on this property, we present direct-coupled architectures in which level-1 suppliers are attached directly to the level-2 core. D1 consumes each T state in its supplier output slot and removes the interface entirely; D2 adds one small buffer per supplier so that the supplier continues distilling without pause. Because the method changes only how a supplied T state is consumed, it applies to the many factories that perform $\pi/8$ rotations by multi-patch Pauli measurements. As an example, we apply it to a CCZ factory and compare it with Litinski's factory under the same logical-error model and the same area accounting. For a maximum output period of 120 cycles, D1 and D2 reduce the area by 24\% and 29\% without increasing the space-time volume, and the smallest D1 factory is 37\% smaller than the smallest Litinski design. The same approach also reduces the area of two-level T factories.
\end{abstract}


\section{Introduction}\label{sec:intro}
Surface-code quantum computers cannot perform universal computation with Clifford operations alone, so the magic states required for non-Clifford operations are produced separately and supplied to the computation~\cite{fowler2012,litinski2019game}. Magic states are obtained by preparing several noisy states and purifying them by distillation~\cite{bravyi2005,bravyi2012}, and for Toffoli-heavy algorithms it is cheaper to distill controlled-controlled-Z (CCZ) states directly than to synthesize the same operation from T states~\cite{jones2013,eastin2013,gidney2019ccz}. In large-scale resource estimates, magic-state factories are among the largest components of the processor~\cite{gidney2021rsa,beverland2022,gidney2025rsa}, and the qubits they occupy cannot be used for algorithmic data. Factory area therefore matters most where the qubit count, rather than run time, is the limit: early fault-tolerant devices, chips or modules of fixed size, and architectures that place several factories next to the data they serve.

Among surface-code CCZ factories, two are commonly used as references. The factory of Gidney and Fowler (GF)~\cite{gidney2019ccz} reaches an output error of $5.3\times10^{-11}$ at physical error rate $p=10^{-3}$ and outputs one CCZ state every 170.5 cycles with about 138,000 qubits. Litinski~\cite{litinski2019msd} reduced this cost substantially by assigning separate X, Z, and temporal distances to each patch; his two-level factory outputs every 60 cycles with about 47,000 qubits. We use GF as the reference for output quality, and Litinski's factory, which is built from the multi-patch measurements that our method modifies, as the concrete baseline for comparison, called \emph{L}.

Rotations by multi-patch Pauli measurements, introduced in the game of surface codes~\cite{litinski2019game}, are the common building block of lattice-surgery magic-state factories. Besides Litinski's factories~\cite{litinski2019msd}, they are used in factories with twist-free and temporally encoded lattice surgery~\cite{chamberland2022,prabhu2022} and with low-cost joint measurements~\cite{kang2024}, and factory layouts of this kind underlie resource estimates and error studies~\cite{beverland2022,chadwick2024}. A rotation is applied by a multi-patch Pauli measurement between data patches and a T-state patch, and the $\pi/4$ correction required by the measurement outcome is handled by an additional $\ket{0}$ patch (the auto-corrected rotation). A two-level factory adds transport channels that carry each level-1 output out of its supplier and waiting spots ($\ket{m}$ spots) where the transported states stay until consumption. We call these three components the \emph{interface}. In Litinski's area formula the interface occupies 30\% of the published factory and 33\% of the re-optimized design used below; a substantial part of the factory is spent not on distillation but on the connection between its levels.

We propose a way to remove this connection area. The key ingredient is \textbf{adaptive X/Y consumption} (Section~\ref{sec:adaptive}), an application of the adaptive-measurement principle of gate teleportation to the inter-level interface of magic-state factories. If the T-state patch used for a $\pi/8$ rotation is read out in the X or Y basis according to the merge outcome, the basis choice alone performs the $\pi/4$ correction, because a Y readout equals an X readout preceded by an $S^\dagger$ gate. The $\ket{0}$ correction patch and its additional merge are therefore unnecessary. We prove the selection rule for both rotation signs and, combined with the in-place Y readout of Gidney~\cite{gidney2024y}, complete each rotation at the position of the T-state patch.

With adaptive consumption, there is no longer a reason to move T states next to the core. Using this property, we present \textbf{direct-coupled architectures} that remove the interface (Section~\ref{sec:arch}). D1 attaches the output slot of each level-1 supplier directly to the merge ancilla of the level-2 core and consumes the T state where it was produced; with no transport channel, waiting spot, or correction patch, it is the smallest factory. D2 places one buffer next to each supplier, so that after handing over its T state the supplier immediately continues with the next distillation; it delivers a higher throughput with a small area.

The method is not tied to a particular factory. It only requires that $\pi/8$ rotations are performed by multi-patch measurements and that the supplier exposes the boundary of its output patch at the block edge, so it can be applied to the factories above and, more generally, to any factory built from multi-patch measurements. We use L as a concrete example and give designs and resources for a CCZ factory built from 15-to-1 suppliers and an 8-to-CCZ core; two-level T factories serve as a second example (Section~\ref{sec:tfactory}). Compared with L re-optimized under the same error model and the same area accounting, the direct-coupled factories are 24--29\% smaller at the same output-period limit without increasing the space-time volume. D1 is also the smallest CCZ factory overall, and under a fixed qubit budget the direct-coupled factories produce more CCZ states. Similar savings can be expected in other factories that use the same interface.

Other work reduces the cost of magic states at different points of the pipeline. Zero-level distillation~\cite{itogawa2024} lowers the error of the raw states, magic-state cultivation~\cite{gidney2024cultivation} grows high-fidelity T states within a surface-code patch, and recent large-scale estimates combine cultivated states with a small distillation stage~\cite{gidney2025rsa}. Our work is complementary: it addresses the connection area between distillation levels, and because direct coupling only requires a T state at the boundary of the core's merge ancilla, it can be combined with these methods.

Section~\ref{sec:prelim} reviews $\pi/8$ rotations by multi-patch measurements, the CCZ circuit, and the supplier. Section~\ref{sec:adaptive} presents adaptive X/Y consumption, and Section~\ref{sec:arch} the layouts and schedule of D1 and D2. Section~\ref{sec:results} evaluates the resources, and Section~\ref{sec:conclusion} concludes.

\section{Preliminaries}\label{sec:prelim}
\begin{figure}[t]\centering
\includegraphics{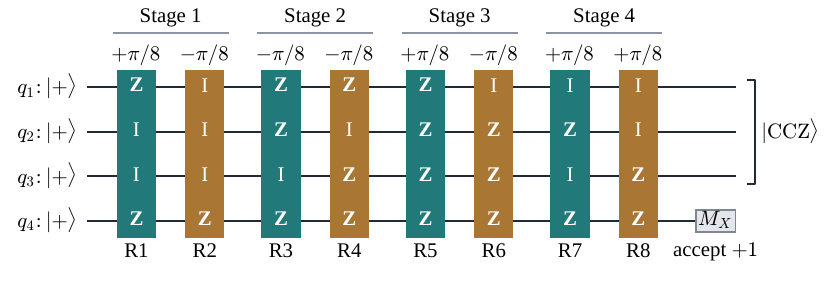}
\caption{Eight-rotation CCZ distillation circuit executed in four stages. Each column R1--R8 is one rotation with the indicated angle and Pauli operator (Z or I on each qubit). Teal rotations run through the upper ancilla row and ochre rotations through the lower row, so two rotations proceed in parallel in every stage. Qubit 4 is finally measured in the X basis and the batch is accepted on $+1$.}\label{fig:circuit}
\end{figure}

\subsection{$\pi/8$ Rotations by Multi-Patch Measurements}\label{sec:gosc}
A rotation $U=e^{is\pi P/8}$ ($s=\pm1$) about a $Z$-type Pauli operator $P$ is implemented by consuming one T state $\ket{A}=(\ket{0}+e^{i\pi/4}\ket{1})/\sqrt2$~\cite{litinski2019game}. First, $P\otimes Z_A$ between the data patches and the T-state patch is measured by lattice surgery~\cite{horsman2012}, with outcome $m=\pm1$; the measurement repeats syndrome extraction for $d_m$ cycles in an ancilla region between the patches. If the T-state patch is then read out in the X basis, the result is either the desired rotation or one that differs by $\pi/4$, depending on $m$; in the latter case a $\pi/4$ correction $e^{\pm i\pi P/4}$ is required.

Litinski's auto-corrected rotation~\cite{litinski2019game} uses one more $\ket{0}$ patch for this correction. It performs one more measurement between the T-state patch and the $\ket{0}$ patch and frees the T-state patch at once by an X readout; after $m$ is decoded, the readout basis of the $\ket{0}$ patch completes the correction. The T-state position is thus free while the decoder works, and the waiting is done by the $\ket{0}$ patch, at the cost of a correction patch and an additional merge region for every rotation.

\subsection{Eight-Rotation CCZ Circuit}\label{sec:ccz}
CCZ states are produced by a distillation circuit based on the $[[8,3,2]]$ code~\cite{litinski2019msd,gidney2019ccz}; we use Litinski's four-qubit form~\cite{litinski2019msd}. Logical qubits 1--4 are prepared in $\ket{+}$, and for every $v=(v_1,v_2,v_3)\in\{0,1\}^3$ one rotation
\begin{equation}
P_v=Z_1^{v_1}Z_2^{v_2}Z_3^{v_3}Z_4,\quad U_v=e^{\,is_v\pi P_v/8},\quad s_v=-(-1)^{|v|},\label{eq:rot}
\end{equation}
is applied, eight in total. Here $s_v$ is the sign of the rotation angle ($\pm\pi/8$) and $|v|$ is the number of ones in $v$; flipping all signs at once gives the same product. Since $\sum_v(-1)^{|v|}Z_1^{v_1}Z_2^{v_2}Z_3^{v_3}=\prod_{i=1}^3(I-Z_i)=8\ket{111}\bra{111}$, the product of the eight commuting rotations is
\begin{equation}
\prod_vU_v=\exp\!\big(-i\pi\ket{111}\bra{111}\otimes Z_4\big)=\mathrm{CCZ}_{123}\otimes I_4.\label{eq:ccz}
\end{equation}
Finally, qubit 4 is measured in the X basis; the batch is accepted on outcome $+1$, and qubits 1--3 hold $\ket{CCZ}$. Each rotation consumes one supplied T state, so one CCZ state needs eight.

The core has one merge-ancilla row above and one below, and each row performs one measurement at a time. Because every rotation involves qubit 4, at most two rotations, one per row, can run simultaneously. The eight rotations are therefore executed in pairs over four stages (Fig.~\ref{fig:circuit}); since they commute, any pairing gives the same result.

\subsection{Suppliers}\label{sec:supplier}
A level-1 supplier is a block that distills T states. Direct coupling requires only that the $Z$ boundary of its output patch faces the block edge, with a free ancilla region in between. We use Litinski's 15-to-1 block~\cite{litinski2019msd} unchanged for all factories: the distilled state remains in an output qubit at one end of the block, next to an ancilla region that extends to the block edge and touches the $Z$ boundary of the output patch. We call the output patch together with this region the \emph{output slot}. Other 15-to-1 implementations with this property, e.g., one using four-qubit $ZY$ measurements~\cite{kang2024}, can also serve as suppliers.

\section{Adaptive X/Y Consumption}\label{sec:adaptive}
\begin{figure}[t]\centering
\includegraphics{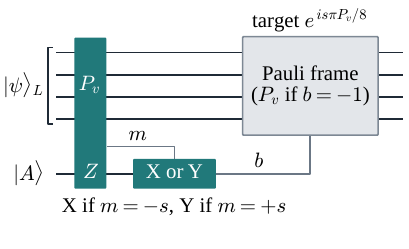}
\caption{Adaptive X/Y consumption. $P_v\otimes Z_A$ between the logical state $\ket{\psi}_L$ and the T state $\ket{A}$ is measured with outcome $m$. Once $m$ is decoded, it is compared with the target sign $s$ and the T-state patch is read out in X or Y. If the readout outcome is $b=-1$, $P_v$ is recorded in the Pauli frame.}\label{fig:adaptive}
\end{figure}

The $\pi/4$ correction of Section~\ref{sec:gosc} is needed because the T-state patch is always read out in the X basis. A Y-basis readout equals an X-basis readout preceded by an $S^\dagger$ gate ($SXS^\dagger=Y$), which is exactly the required $\pi/4$ correction. Choosing the readout basis from $m$ therefore removes the need for a correction patch (Fig.~\ref{fig:adaptive}):

\emph{Proposition:} After measuring $P\otimes Z_A$ with outcome $m$, read out the T-state patch in the X basis if $m=-s$ and in the Y basis if $m=+s$. With readout outcome $b=\pm1$, the operation applied to the data is, up to global phase and normalization, $P^{(1-b)/2}e^{is\pi P/8}$: the target rotation is applied exactly, and only for $b=-1$ an additional Pauli operator $P$ remains.

\emph{Proof:} Decompose the data state into eigencomponents of $P$, $\ket{\psi}=\ket{\psi_+}+\ket{\psi_-}$ with $P\ket{\psi_\lambda}=\lambda\ket{\psi_\lambda}$. Measuring $P\otimes Z_A=m$ on $\ket{\psi}\ket{A}$ pairs the component with $\lambda=m$ with $\ket{0}_A$ and the component with $\lambda=-m$ with $e^{i\pi/4}\ket{1}_A$:
\begin{equation}
\ket{\psi_m}\ket{0}+e^{i\pi/4}\ket{\psi_{-m}}\ket{1}.\label{eq:post}
\end{equation}
Projecting onto the X eigenstate $(\ket{0}+b\ket{1})/\sqrt2$ leaves $\ket{\psi_m}+b\,e^{i\pi/4}\ket{\psi_{-m}}$ on the data, and projecting onto the Y eigenstate $(\ket{0}+ib\ket{1})/\sqrt2$ leaves $\ket{\psi_m}+b\,e^{-i\pi/4}\ket{\psi_{-m}}$. Since
\begin{equation}
e^{i\theta P}\ket{\psi}=e^{i\theta m}\big(\ket{\psi_m}+e^{-2i\theta m}\ket{\psi_{-m}}\big),\label{eq:phase}
\end{equation}
the relative phases $e^{+i\pi/4}$ and $e^{-i\pi/4}$ correspond to $\theta=-m\pi/8$ (X readout) and $\theta=+m\pi/8$ (Y readout). Hence the X readout gives the target rotation $\theta=s\pi/8$ when $m=-s$, and the Y readout when $m=+s$; the opposite basis gives $-s\pi/8$. For $b=-1$ only the sign of the $\ket{\psi_{-m}}$ component changes, which is one more application of $P$.

The remaining $P$ is $Z$-type and commutes with all subsequent rotations of the CCZ circuit, so it is recorded in the Pauli frame rather than applied; it only affects the final X measurement of qubit~4 and a $Z$ correction of the outputs, both handled classically. We also computed all $4^8$ measurement branches of the eight-rotation circuit by state-vector simulation and confirmed that the acceptance probability and fidelity after Pauli-frame correction are one (Appendix~\ref{app:circuits}).

\emph{Relation to the auto-corrected rotation:} Choosing the readout basis from a measurement outcome is a known principle of gate teleportation, and we do not claim it as a new primitive. Our contribution is to use it inside magic-state factories, where it changes where the $\pi/4$ correction takes place. The auto-corrected rotation needs a $\ket{0}$ patch and an additional merge for every rotation and defers the correction to that patch, whereas adaptive consumption places it in the readout basis of the T-state patch itself. A rotation then needs only the T-state patch and one merge, so a T state can be consumed wherever it shares a boundary with the core; the architectures of Section~\ref{sec:arch} rely on this property. The rule does not depend on the circuit and applies to any factory built from multi-patch measurements. It also does not depend on the surface code: it requires only a joint Pauli measurement and an X or Y readout of the T-state qubit, whereas the direct-coupled layouts below are specific to planar surface-code patches.

\emph{Readout time:} An X readout takes one cycle. For the Y readout we use Gidney's in-place construction~\cite{gidney2024y}, which preserves the distance by using twist defects within the bounding box of the patch and takes
\begin{equation}
t_Y(d)=\lfloor d/2\rfloor+2\ \text{cycles}\label{eq:ty}
\end{equation}
for a patch of distance $d$. Since the readout follows the merge with the core, the orientation of the patch is unconstrained. Gidney notes that additional waiting may be needed when the twist of the readout must be separated from other twist defects; all merges of our factory measure $Z$-type operators only and use no twist defects, so (\ref{eq:ty}) is used as the readout time. The logical error of the Y readout is treated in Appendix~\ref{app:model}.

\emph{Order of decoding and readout:} Because the readout basis depends on $m$, the T-state patch is read out after $m$ is decoded. Decoding takes the reaction time $R$, for which we use $R=10$ cycles~\cite{fowler2012time}. The waiting formerly done by the correction patch is thus done by the T-state patch itself, and where this waiting takes place distinguishes the two architectures of Section~\ref{sec:arch}.

\section{Direct-Coupled Architectures}\label{sec:arch}
The design principle of the direct-coupled architectures is to \emph{consume a T state where it is produced}. With adaptive consumption a T state can be consumed as soon as it shares a boundary with the core, so attaching the suppliers directly to the core makes the connection between them unnecessary. Table~\ref{tab:compare} shows how this principle changes the interface. L uses transport channels, waiting spots, and correction patches to bring T states next to the core; D1 removes all three, and D2 keeps only a small buffer next to each supplier.

We denote by $d_X$ and $d_Z$ the spatial distances against X- and Z-type logical errors and by $d_m$ the number of cycles of each lattice-surgery measurement; choosing spatial and temporal distances separately is the key to small factories~\cite{litinski2019msd}. Level-2 (core) values carry a subscript 2, buffers have size $d_{p2}$, and $d_{p1}$ is the size of the small ancilla strips of the supplier.

\begin{table}[t]\centering
\caption{Interface of the three factories for the 120-cycle designs of Table~\ref{tab:main} (four suppliers). A dash means that the component is absent.}\label{tab:compare}
\setlength{\tabcolsep}{4pt}\footnotesize
\begin{tabular}{lccc}\toprule
 & L & D1 & D2\\\midrule
Transport channels & 4 & -- & --\\
Waiting spots / buffers & 4 & -- & 4\\
$\ket{0}$ correction patches & 2 & -- & --\\
Interface area (qubits) & 12,600 (33\%) & 0 & 1,352 (5\%)\\\bottomrule
\end{tabular}
\end{table}

\begin{figure}[t]\centering
\includegraphics[width=\linewidth]{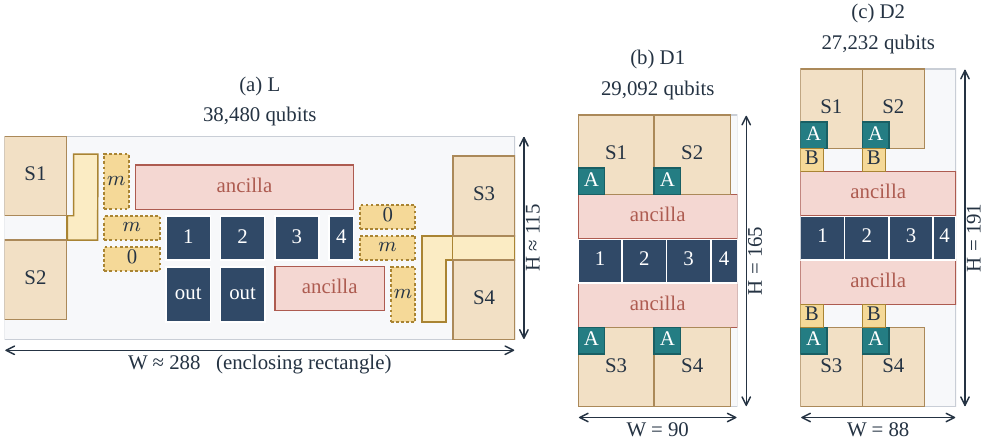}
\caption{Layouts of the three factories selected under a 120-cycle limit, to scale in data-qubit sites. (a) L: Litinski's interface. Suppliers S1--S4 lie on both sides, and transport channels, waiting spots $\ket{m}$ (labeled $m$), and correction patches $\ket{0}$ (labeled 0) surround the core (1--4) and its ancilla (pink). (b) D1: the output slot A of each supplier touches an ancilla row (pink) of the core directly. (c) D2: one buffer B next to each output slot A. Suppliers are beige, the core navy, and interface components and buffers yellow. Supplier distances are optimized for each factory: $(15,5,7)$ for L and D2 and $(15,7,5)$ for D1 (Section~\ref{sec:footprint}). The light rectangles enclose the layouts; for all three factories only the components are counted in the area.}\label{fig:arch}
\end{figure}

\begin{figure}[t]\centering
\includegraphics{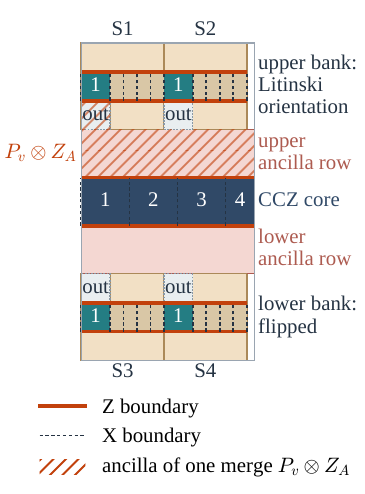}
\caption{Boundary arrangement of D1. Qubit 1 of each supplier holds the distilled T state next to its \emph{out} region. The upper suppliers keep their orientation and the lower suppliers are flipped upside down, so that the merge region (hatched) touches only $Z$ boundaries of the T state and the core patches.}\label{fig:boundary}
\end{figure}

\subsection{D1: In-Place Consumption}
The idea of D1 is not to move T states at all: the output slot of each supplier serves directly as one end of a core merge.

The layout is shown in Fig.~\ref{fig:arch}(b). Core patches 1--4 lie in a row in the middle, with one merge-ancilla row above and one below, each spanning the full width of the supplier banks. Half of the suppliers are placed above the upper row and half below the lower row. The upper suppliers keep Litinski's orientation so that the \emph{out} region of each output slot touches the upper ancilla row; the lower suppliers are flipped upside down so that their \emph{out} regions face up and touch the lower row. This is only a change of layout orientation; flipping keeps the boundary types, so in both banks the boundary facing the ancilla row is the $Z$ boundary of the output patch, as is the boundary of each core patch facing that row (Fig.~\ref{fig:boundary}). The measurement of $P_v\otimes Z_A$ between the core and a T state therefore uses the \emph{out} region and the ancilla row as one ancilla.

After the merge, the T-state patch waits in place for $m$ to be decoded and is read out in X or Y by the rule of Section~\ref{sec:adaptive}. No transport channel, waiting spot, correction patch, or buffer is needed, so a single output slot serves both production and consumption. D1 is consequently the smallest of the three layouts.

\subsection{D2: One Buffer per Supplier}
D2 keeps the direct-coupled structure of D1 and places one buffer per supplier between the supplier bank and the ancilla row (Fig.~\ref{fig:arch}(c)). The buffer separates the place where a T state is \emph{produced} from the place where it \emph{waits to be consumed}.

\begin{figure}[t]\centering
\includegraphics[width=\linewidth]{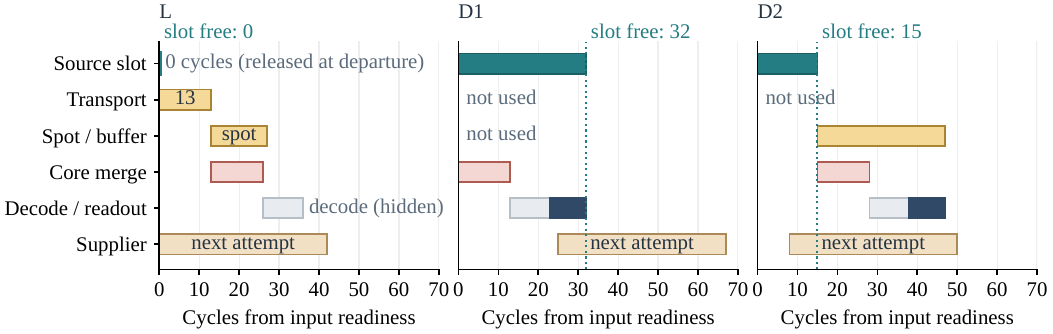}
\caption{Sequence for one supplied T state in L, D1, and D2. Rows show the occupation of the supplier output slot, transport, waiting spot or buffer, core merge, decoding and readout, and the next distillation of the supplier. Dotted lines mark the release of the output slot; in L the slot is released at departure (0 cycles), as assumed in Section~\ref{sec:dims}. L consumes the state through transport channel, waiting spot, and correction patch; D1 consumes it directly from the output slot; D2 consumes it after handing it to the buffer. Example values: $d_X=d_{p2}=15$, $d_m=7$, $d_{m2}=13$, $R=10$, Y readout.}\label{fig:timeline}
\end{figure}

When distillation finishes, the T state is moved from the output slot to the buffer by lattice surgery, which takes $h=\max(d_X,d_{p2})$ cycles. The supplier can then run its next distillation to completion, while merge, decoding, and readout take place in the buffer. In the example of Fig.~\ref{fig:timeline}, D2 returns the output slot to the supplier after 15 cycles. Because the supplier keeps distilling, the same number of suppliers reaches a shorter output period.

The buffer plays the role of L's waiting spot, but D2 has no transport channel and no correction patch: the buffer sits next to the supplier, so nothing has to be carried far, and adaptive consumption removes the correction patch. Its interface area is therefore about one tenth of that of L (Table~\ref{tab:compare}). D1 minimizes the area itself, whereas D2 raises the throughput while keeping the area small; the two designs take different points between area and speed on the same principle.

\subsection{Area and Schedule}\label{sec:dims}
The areas of D1 and D2 are computed in the same way as for L, as a sum of component areas. The components are $n$ supplier blocks, the core row (three output patches and one syndrome patch), two merge-ancilla rows, and, for D2, the buffers. Since the merge-ancilla rows must reach every supplier, their length is the larger of the core row and a bank of $n/2$ suppliers,
\begin{equation}
W=\max\!\big(3d_{X2}+d_{Z2},\ \tfrac n2(d_X+4d_Z)\big).\label{eq:w}
\end{equation}
With $A_s=(d_X+4d_Z)\,3d_X+2d_{p1}$ the area of one supplier block including its small $\ket{+}$ ancilla strips, counted exactly as for L in (\ref{eq:ql}), the physical qubit counts including measurement qubits are
\begin{equation}
Q_{D1}=2\big[nA_s+(3d_{X2}+d_{Z2})\,d_{X2}+2d_{X2}W\big]\label{eq:qd1}
\end{equation}
and $Q_{D2}=Q_{D1}+2nd_{p2}^2$, which adds the buffers. For the designs of Fig.~\ref{fig:arch}, $W=90$ (D1) and $88$ (D2), giving 29,092 and 27,232 qubits. When the supplier banks set the width, the core row has spare width, and the syndrome patch can be widened at no extra area to reduce its error.

A CCZ state is produced in the four stages of Section~\ref{sec:ccz}. In each stage both ancilla rows receive one T state each, and the two merges start together once both states and both rows are ready. A merge takes $d_{m2}$ cycles, decoding of its outcome $R$ cycles, and the readout one cycle (X) or $t_Y$ cycles (Y). A level-1 distillation takes $6d_m$ cycles and is repeated upon rejection.

The supplier reuse rules are as follows. The first of the six steps of Litinski's block does not use the output qubit, so a new distillation may start up to $d_m$ cycles before the output slot is freed. In D1 the output slot is freed when the readout finishes; in D2 it is freed when the transfer to the buffer finishes, and the buffer stays reserved until its readout ends. L frees the output slot when transport departs. This assumes that the slot of L is free as soon as transport starts, which favors L because a lattice-surgery move also occupies the source patch until it completes. Appendix~\ref{app:noise} evaluates the factories with a common release rule, under which D2 matches or improves on L in space-time volume. In the reference policy, the next CCZ state is started when all four stages of the previous one have been read out.

\section{Resource Evaluation}\label{sec:results}
We compare L, D1, and D2 under identical conditions. The logical error model is Litinski's model extended by waiting and readout errors (Appendix~\ref{app:model}). The reference conditions are $p=10^{-3}$, an output-error target of $5.3\times10^{-11}$, and $R=10$ cycles. The target is the output error reported for the GF factory at $p=10^{-3}$~\cite{gidney2019ccz}, so every factory delivers CCZ states of at least the quality of this reference. It is a constraint rather than a value to be matched: because distances change in odd steps, the selected designs have output errors at or below the target, and their actual values are listed in Table~\ref{tab:dist}. For each of the three factories, all distances of both levels, including those of the suppliers, the buffer size, and the supplier count $n\in\{2,4,\dots,12\}$ are searched independently, and the design with the fewest qubits that meets the target and the period limit is selected; L is thus not taken with its original values but re-optimized under the same model. GF is included with its published values. The area of L follows Litinski's component-area formula~\cite{litinski2019msd,litinskicode} generalized to $n$ suppliers, (\ref{eq:ql}), and the areas of D1 and D2 follow (\ref{eq:qd1}) and its D2 extension, which count components in the same way.

The \emph{output period} is the mean number of cycles per accepted CCZ state, evaluated by stochastic simulation with random supplier rejection and random X/Y choices, and the \emph{space-time volume} is the product of area and output period. Fixing a maximum output period compares areas, and fixing a qubit budget compares CCZ rates. As the representative operating point we use a maximum output period of 120 cycles. This lies between the output periods of the two reference factories, the published Litinski factory (60 cycles) and GF (170.5 cycles), and corresponds to a moderate rate at which the interface of L is fully in use. The limit is a design parameter that can be adjusted to the rate an application requires; Table~\ref{tab:main} also lists a slower limit of 160 cycles, and Fig.~\ref{fig:fronts} shows the comparison over the full range.

\subsection{Area and CCZ Rate}\label{sec:footprint}
\begin{figure}[t]\centering
\includegraphics{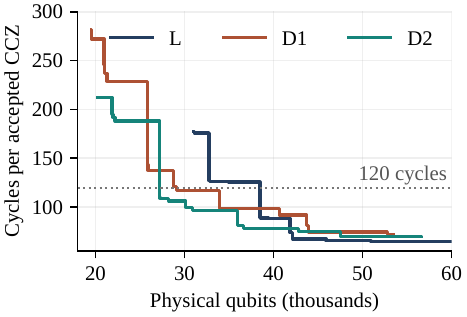}
\caption{Physical qubits versus output period. Each curve gives the shortest output period achievable with at most the given number of qubits.}\label{fig:fronts}
\end{figure}

Fig.~\ref{fig:fronts} shows the shortest output period each factory reaches with a given number of qubits, and Table~\ref{tab:main} lists representative designs.

\begin{table}[t]\centering
\caption{Representative designs at $p=10^{-3}$ and output-error target $5.3\times10^{-11}$. Qubits in thousands; ratios relative to L under the same condition.}\label{tab:main}
\setlength{\tabcolsep}{2.6pt}\footnotesize
\begin{tabular}{llrrrcc}\toprule
condition & factory & $n$ & qubits & cycles & area/L & volume/L\\\midrule
period $\le120$ & L & 4 & 38.5 & 90 & 1.00 & 1.00\\
 & D1 & 4 & 29.1 & 117 & 0.76 & 0.99\\
 & D2 & 4 & 27.2 & 110 & 0.71 & 0.87\\\midrule
period $\le160$ & L & 2 & 32.8 & 127 & 1.00 & 1.00\\
 & D1 & 4 & 25.9 & 143 & 0.79 & 0.89\\
 & D2 & 4 & 27.2 & 110 & 0.83 & 0.72\\\midrule
min.\ area & L & 2 & 31.1 & 177 & 1.00 & --\\
 & D1 & 2 & 19.5 & 282 & 0.63 & --\\
 & D2 & 2 & 20.2 & 212 & 0.65 & --\\\midrule
published & GF~\cite{gidney2019ccz} & -- & 138.4 & 171 & -- & --\\
 & Litinski~\cite{litinski2019msd} & 6 & 47.0 & 60 & -- & --\\\bottomrule
\end{tabular}
\end{table}

In Fig.~\ref{fig:fronts} the curves of the direct-coupled factories lie to the left of L, i.e., at fewer qubits, over a wide range of output rates. Under the 120-cycle limit, D1 and D2 are 24\% and 29\% smaller than L, and their space-time volume is equal to that of L (D1) or 13\% smaller (D2). Since 33\% of the L area is interface, most of this difference comes from removing the connection: the effect of eliminating the correction patches by adaptive consumption and attaching the suppliers directly to the core appears directly in the area.

Because the supplier distances are part of the optimization, the three factories in Fig.~\ref{fig:arch} do not use identical suppliers: under the 120-cycle limit, L and D2 use $(15,5,7)$ suppliers, whereas D1 uses $(15,7,5)$ suppliers, which distill faster ($d_m=5$) but are wider ($d_Z=7$). The reason is the occupancy of the output slot. In D1 the slot stays occupied until merge, decoding, and readout of its T state are finished, so reaching 120 cycles requires suppliers that distill faster. In D2 the buffer frees the supplier right after the handoff, so the smaller suppliers already reach this rate. This is why D2 is slightly smaller than D1 under the 120-cycle limit: its suppliers occupy about 2,900 fewer qubits, which outweighs the 1,352 qubits of its buffers.

When the suppliers are identical, the ordering follows directly from the interface. Among the validated designs with $(15,5,7)$ suppliers, D1 needs 25,912 qubits at 138 cycles, D2 27,232 qubits at 110 cycles, and L 38,480 qubits at 90 cycles: D1 is the smallest and D2 is faster. Accordingly, once the period limit allows D1 to use the smaller suppliers, D1 becomes the smallest factory, as under the 160-cycle limit in Table~\ref{tab:main} (21\% smaller than L), and without a period limit D1 is 37\% smaller than the smallest L design. D1 thus minimizes the area itself, while D2 delivers a fast output rate with a small area; in either case a direct-coupled factory is smaller than L.

When the qubit count is fixed by the chip size or by the space left after placing the data, the same area difference becomes a difference in CCZ rate. Packing as many copies of a design as fit into a budget, direct-coupled factories fit into small budgets where no L factory fits, and for budgets holding one or two factories they produce more CCZ states over several budget ranges. The largest gain occurs at a budget of 32,000 qubits, where the best L, D1, and D2 factories produce 5.7, 8.6, and 10.3 CCZ states per 1000 cycles, a factor of 1.8 for D2 over L.

\subsection{Results Under Various Conditions}\label{sec:sens}
To check that these results do not rely on particular assumptions, we re-optimized all three factories from scratch and validated them stochastically for each changed condition: an additional in-place Y-readout fault, twice slower decoding, a margin on the output-error target, two measured noise anchors from Stim, a lower physical error rate ($p=10^{-4}$, with an output-error target of $7.3\times10^{-14}$), and merge-ancilla rows of twice the height (Appendix~\ref{app:noise}). In all cases, D1 and D2 had a smaller area than L under the 120-cycle limit. D2 in particular remained at least 23\% smaller than L under every condition at $p=10^{-3}$, showing that separating supplier and consumption by one buffer is robust against these changes. The height of the merge-ancilla rows can be kept at its reference value with hook-aware syndrome extraction~\cite{hirai2026hooks}.

\subsection{Application to T Factories}\label{sec:tfactory}
Adaptive consumption and direct coupling are not limited to the CCZ circuit. Each level-2 rotation of a two-level 15-to-1 factory also consumes one level-1 T state by a $Z$-type measurement, so adaptive consumption applies unchanged. Litinski's two-level (15-to-1)$\times$(15-to-1) T factories~\cite{litinski2019msd} use the same interface as the CCZ factory, which occupies 26--28\% of their area. For all four published T factories, attaching the suppliers of the same distances directly to the core as in D1 reduces the area, computed with the same formula, by 18--26\%. This shows that the method applies not to one particular factory but to multi-patch-measurement factories in general~\cite{litinski2019game}.

\section{Conclusion}\label{sec:conclusion}
We applied the adaptive-measurement principle of gate teleportation to the $\pi/8$ rotations of magic-state factories as adaptive X/Y consumption. It handles the $\pi/4$ correction not with a separate correction patch but by the choice of readout basis of the T-state patch itself, and we proved the selection rule for both rotation signs. A T state thereby becomes a resource that can be consumed as soon as it shares a boundary with the core, and the transport channels, waiting spots, and correction patches that connected level~1 to level~2 are no longer needed.

On this property we built direct-coupled architectures that consume a T state where it is produced. D1 uses the supplier output slot as one end of a core merge and removes the connection entirely; D2 places one buffer per supplier so that the supplier keeps distilling. Compared with Litinski's factory under the same model and the same area accounting, the direct-coupled CCZ factories reduce the area by 24--29\% without increasing the space-time volume, and this holds under a range of noise and operating conditions. The same approach applies to T factories. Adaptive consumption and direct coupling can therefore serve as a general way to make magic-state factories small on fault-tolerant devices with a limited number of qubits.

Because the method is not tied to the type of supplier, it may in the future also be applied to factories supplied with cultivated T states or to factories based on other distillation protocols.

\appendix
\section{Logical Error Model and Search}\label{app:model}
\emph{Reference error rates:} We follow the channel sequences of Litinski's five-qubit level-1 and four-qubit level-2 calculations~\cite{litinski2019msd,litinskicode}: Pauli and $\pm\pi/4$ rotation-error channels are applied to the distillation qubits, followed by postselection. The logical error rate per code cycle of a $d\times d$ patch is
\begin{equation}
p_L(p,d)=0.1(100p)^{(d+1)/2},\quad p_s=Sp_L,\quad p_t=Tp_L,\label{eq:pl}
\end{equation}
where $S$ and $T$ are space-like and time-like multipliers, $S=T=1$ in the reference model. For a patch of area $\area$ (in data-qubit sites) stored at distance $d$ for $t$ cycles, we approximate the error probability per error type (X or Z) by $\area\,p_s(d)\,t/(2d^2)$, and for a merge over an ancilla of area $\area$ lasting $d_m$ cycles, the outcome-error probability by $\area\,p_t(d_m)/(2d_m)$; both reduce to Litinski's expressions for $S=T=1$. The merge ancilla is $W+d_X$ long in D1 and $W+d_{p2}$ in D2.

\emph{Waiting errors:} For each supplied state the scheduler records the mean waiting time $\tau_s$ in the supplier output slot, the mean residency $\tau_b$ in a D2 buffer, and the transfer time $h$. The additional input errors are
\begin{align}
e_{D1}&=\tfrac12\tau_sp_s(d_X),\label{eq:ed1}\\
e_{D2}&=\tfrac12\tau_sp_s(d_X)+\tfrac12\tau_bp_s(d_{p2})\notag\\
&\quad+\tfrac12h\,p_s(\min\{d_X,d_{p2}\}),\label{eq:ed2}
\end{align}
where the last term charges the transfer at the smaller of the two patch sizes. The waiting times include the readout.

\emph{Error propagation under adaptive consumption:} With adaptive consumption, an error of the T-state patch propagates differently in the two readout branches. A $Z$ error, before or after the merge, becomes the Pauli error $P$ in both the X and Y readouts. An $X$ error before the merge becomes a rotation error $e^{-is\pi P/4}$ in the X branch and $e^{+is\pi P/4}$ in the Y branch; since the two branches are equally likely, the average is
\begin{align}
&\tfrac12\big(e^{-is\pi P/4}\rho\,e^{+is\pi P/4}+e^{+is\pi P/4}\rho\,e^{-is\pi P/4}\big)\notag\\
&\qquad=\tfrac12\big(\rho+P\rho P\big).\label{eq:avg}
\end{align}
An $X$ error after the merge leaves the X readout unchanged and flips the Y readout, which again gives $P$ with probability one half. For waiting-error probabilities $q_X$ and $q_Z$, the effective probability of $P$ is therefore $q_Z+q_X/2$, and the direct-coupled factories are evaluated with this channel. A wrong merge outcome $m$ selects the opposite basis and gives a rotation error of fixed sign, which is treated as in Litinski's model.

\emph{Y-readout fault:} The outcome of the in-place Y readout is affected by both X and Z logical errors of the patch, and Gidney reports a Y-basis logical error rate 1.5--2.5 times the sum of the X and Z memory error rates~\cite{gidney2024y}. For the analysis of Section~\ref{sec:sens} we apply the upper end of this factor and add $e_Y=2.5\,p_s(d)\,t_Y(d)$ for each Y readout as an estimate. This error propagates through the CCZ circuit and postselection along the same path as a $Z$ error of the supplied T state.

\emph{Treatment of L:} For L we keep Litinski's transport-path error and his treatment of the correction patch and add the waiting beyond two merge intervals. The area of L is the sum of component areas,
\begin{align}
Q_L=2\Big[&(3d_{X2}+d_{Z2})\,3d_{X2}\notag\\
&+n\{(d_X+4d_Z)(3d_X+d_{p2}/2)+2d_{p1}\}\notag\\
&+20d_{p2}^2+2d_{X2}d_{p2}\Big],\label{eq:ql}
\end{align}
where the first term is the level-2 core including its two ancilla rows, the second the $n$ suppliers including transport channels and ancilla strips, and the last group the intermediate region with waiting spots and correction patches. The areas of D1 and D2 in (\ref{eq:qd1}) likewise add only component areas.

\emph{Distillation calculation:} For numerical stability we propagate the difference $\Delta\rho=\rho-\rho_0$ from the ideal density matrix $\rho_0$. With $\Pi$ the projector onto accepted outcomes, $F$ the projector onto the ideal accepted output, $a=\mathrm{Tr}(\Pi\Delta\rho)$, and $f=\mathrm{Tr}(F\Delta\rho)$, the rejection probability and output error are
\begin{equation}
p_{\mathrm{fail}}=-a,\qquad \pout=\frac{a-f}{1+a}.\label{eq:pout}
\end{equation}
This calculation was checked against independent 60-digit computations and against Litinski's implementation. For independent $Z$ errors of probability $p_1$ on the eight supplied states, the probability that an accepted output is wrong is $28p_1^2+O(p_1^3)$.

\emph{Search and validation:} At $p=10^{-3}$ we search all odd distances with $d_X$ 9--23 and $d_m$ 3--15 at level~1, $d_{X2}$ 19--33 and $d_{m2}$ 9--25 at level~2, buffer sizes $d_{p2}$ 9--23, and all even supplier counts from 2 to 12. For each qubit count the fastest design is kept, which forms the curves of Fig.~\ref{fig:fronts}. The kept designs are re-evaluated by stochastic simulation (eight seeds with 4096 CCZ batches each), and only designs that still meet the target after every mean waiting time is increased by three standard errors are used; this margin checks the numerical stability of the mean-waiting-time estimate. Table~\ref{tab:dist} lists the distances of the selected designs.

\begin{table}[t]\centering
\caption{Distances of the selected designs: $(d_X,d_Z,d_m);d_{p1}$ at level 1 and $(d_{X2},d_{Z2},d_{m2});d_{p2}$ at level 2. Output error in units of $10^{-11}$.}\label{tab:dist}
\setlength{\tabcolsep}{2.2pt}\footnotesize
\begin{tabular}{llccrrr}\toprule
condition & & level 1 & level 2 & qubits & cycles & error\\\midrule
period $\le120$ & L & $(15,5,7);5$ & $(25,13,17);15$ & 38,480 & 89.5 & 3.42\\
 & D1 & $(15,7,5);7$ & $(25,15,13);-$ & 29,092 & 116.9 & 5.13\\
 & D2 & $(15,5,7);5$ & $(25,13,17);13$ & 27,232 & 110.1 & 4.95\\\midrule
period $\le160$ & L & $(15,7,5);5$ & $(25,13,17);15$ & 32,770 & 127.0 & 5.24\\
 & D1 & $(15,5,7);5$ & $(25,13,15);-$ & 25,880 & 143.1 & 3.18\\
 & D2 & $(15,5,7);5$ & $(25,13,17);13$ & 27,232 & 110.1 & 4.95\\\midrule
min.\ area & L & $(15,5,7);5$ & $(25,13,17);15$ & 31,090 & 177.2 & 4.01\\
 & D1 & $(15,5,7);5$ & $(25,13,15);-$ & 19,540 & 281.8 & 4.07\\
 & D2 & $(15,5,7);7$ & $(25,13,17);13$ & 20,232 & 212.2 & 3.79\\\bottomrule
\end{tabular}
\end{table}

\section{Noise Calibration and Scheduling Policies}\label{app:noise}
\emph{Absolute noise anchors:} Circuits generated with Stim~\cite{gidney2021stim} were decoded with PyMatching~\cite{higgott2022pymatching} to obtain space-like and time-like failure rates, using memory experiments of $d\times d$ patches for the space-like rate and stability experiments~\cite{gidney2022stability} of $10\times10$ patches for the time-like rate. Gate, reset, and idle errors have strength $p=10^{-3}$ and measurement outcomes flip with probability $q$. The multipliers obtained at $d=7$ are $(S,T)\approx(1.6,0.7)$ for $q=p$ and $(2.8,5.0)$ for $q=5p$; these values are inserted into (\ref{eq:pl}) for the comparison of Section~\ref{sec:sens}.

\emph{Results per condition:} Under the 120-cycle limit, the area relative to L is 0.74--0.97 for D1 and 0.71--0.76 for D2 over the conditions of Section~\ref{sec:sens} at $p=10^{-3}$, and 0.87 (D1) and 0.91 (D2) at $p=10^{-4}$.

\emph{Scheduling policy:} In the reference policy, the next CCZ state starts after the previous one is finished. Overlapping consecutive CCZ states or changing when output slots are released changes the minimum space-time volume, but the D2/L ratio stays within 0.82--1.01, i.e., D2 is equal to or smaller than L (without separate extra output storage).

\section{Circuit Checks}\label{app:circuits}
\emph{Adaptive X/Y consumption:} For the eight-rotation CCZ circuit we computed all combinations of merge and readout outcomes ($4^8=65{,}536$) by state-vector simulation. After Pauli-frame correction the acceptance probability and fidelity are one in all branches, and choosing the opposite readout basis never gives the target rotation, in agreement with the proposition of Section~\ref{sec:adaptive}.

\emph{Merge circuits:} We built Stim circuits of individual merges between core patches and T-state patches and checked 96 configurations covering the eight rotation supports, two positions of the T-state patch, and two ancilla heights at $d=5$ and 7. In each configuration the target parity is obtained from measured stabilizers, exactly one logical degree of freedom is removed, and no qubit is used twice in any CNOT layer. With twice the ancilla height, the merge distance is preserved even with the standard CNOT order; the doubled-height analysis of Section~\ref{sec:sens} follows this construction.

\emph{Reproducibility:} Code and data for all tables and figures except the schematics of Figs.~\ref{fig:circuit} and~\ref{fig:adaptive} are available at \url{https://doi.org/10.5281/zenodo.24066001}~\cite{kookcode}, whose README maps each table and figure to the script that produces it. All results were regenerated in a clean environment with identical values, and a separate implementation of the scheduling rules reproduced the output periods for 720 combinations. Among 20,000 random designs drawn from the search range, none improved on the selected curves.

\end{document}